\documentclass[12pt]{article}
\usepackage[left=2.5cm,top=2.50cm,right=2.5cm,bottom=2.50cm]{geometry}
\usepackage{mathrsfs}
\usepackage{amsmath,amssymb,latexsym,color,cancel,graphicx,bbm,colortbl}
\usepackage[english]{babel}
\usepackage[utf8]{inputenc}
\usepackage{ragged2e}
\usepackage{cite}
\usepackage{float}
\usepackage{subcaption}
\begin{document}
\date{}

\title{Phase-space dynamics and multicomponent wave-packet structures in the Dunkl anharmonic oscillator}
\author{D. Ojeda-Guill\'en$^{a}$\footnote{{\it E-mail address:} dojedag@ipn.mx} and R. D. Mota$^{b}$} \maketitle

\begin{minipage}{0.9\textwidth}
\small $^{a}$ Escuela Superior de C\'omputo, Instituto Polit\'ecnico Nacional,
Av. Juan de Dios B\'atiz esq. Av. Miguel Oth\'on de Mendiz\'abal, Col. Lindavista,
Alc. Gustavo A. Madero, C.P. 07738, Ciudad de M\'exico, Mexico.\\

\small $^{b}$ Escuela Superior de Ingenier{\'i}a Mec\'anica y El\'ectrica, Unidad Culhuac\'an,
Instituto Polit\'ecnico Nacional, Av. Santa Ana No. 1000, Col. San
Francisco Culhuac\'an, Alc. Coyoac\'an, C.P. 04430, Ciudad de M\'exico, Mexico.\\

\end{minipage}

\begin{abstract}
We study the phase-space dynamics and multicomponent wave-packet structures of the Dunkl anharmonic oscillator (Kerr medium). Starting from the time-evolved state, we derive analytical expressions for the Dunkl-Husimi $Q$-function and a normalized angular Husimi profile. The Dunkl parameter modifies the coherent-state excitation weights and the parity-dependent phase accumulation, producing angular anisotropy and asymmetric phase-space separation. For the half-integer deformation $\mu=0.5$, the dynamics generate asymmetric multicomponent Husimi patterns with unequal peak heights and an early reconstruction at $t=\pi/\lambda$. For the deformation $\mu=1.0$, the Husimi distribution exhibits broadening while retaining its multimodal structure and unitary evolution. At the fundamental nonlinear revival time, the wave packet is reconstructed for the selected parameter ratio. The standard Kerr dynamics are recovered in the limit $\mu\rightarrow 0$.

\vspace{0.5cm}
\end{abstract}

\section{Introduction}

The study of algebraic deformations in quantum mechanics has expanded since Wigner's inquiries into the commutation relations of the harmonic oscillator \cite{wigner1950}. Following Yang's introduction of reflection operators for spatial quantization \cite{yang1951}, Dunkl formulated a differential-difference operator associated with reflection groups \cite{dunkl1989,dunkl2001}. This formalism introduces a parity-dependent term that modifies the physical properties of the models.

Parallel to these developments, operator techniques related to the Wigner-Heisenberg algebra were investigated. Deformed quadratic operators containing reflection terms close the $su(1,1)$ Lie algebra \cite{jayaraman1990,jayaraman1999,rodrigues2003,delimarodrigues2009,carrion2010,rodrigues2013,plyuschay1996,plyuschay1997}. This algebraic consistency allows the treatment of supercoherent states and superconformal quantum mechanics, establishing a framework for parity-deformed formalisms.

In the last decade, the Dunkl derivative was applied to the harmonic oscillator and Coulomb potentials \cite{genest2013,genest2014a,nos2017}. It has also been applied to spatially varying mass systems and higher-dimensional wave equations \cite{ghazouani2019a,quesne2025,raber2025,benarous2025,hamil2025a,lut2025,hocine2025,hamil2025b,salazar2026}. Recently, these deformations were explored in statistical mechanics and quantum optics, including the study of geometric phases and Mandel parameters in parity-deformed systems \cite{moroz2016,hamil2022b,chung2021,nos2026}.

In quantum optics, nonlinear systems generate non-classical states of light \cite{estes1968,caves1981,gerry1987,perelomov1972,wodkiewicz1985,gerry1989,vourdas1990,yurke1986,nos2014,choreno2018,makarov2018,ramos2021,algarni2022,jones2023,bocanegra2024,choreno2021,vega2024,vega2025}. A standard model in this field is the Tana\'s anharmonic oscillator, or Kerr medium \cite{tanas1984}. This system generates squeezed states, exhibits the collapse and revival of the survival probability \cite{eberly1980,kirchmair2013}, and forms multicomponent wave-packet structures \cite{yurkeStoler1986,vlastakis2013,mirrahimi2014}. These structures have applications in quantum information processing and are studied using phase-space representations, such as the Husimi $Q$-function and the Wigner distribution \cite{husimi1940,milburn1986,buzek1992}. These phase-space tools were also extended to characterize coherent states in parity-deformed formalisms \cite{ghazouani2022,sedaghatnia2023}.

Following this direction, the Dunkl formalism was applied to generalize the Tana's anharmonic oscillator \cite{ojeda2026b}. That work established the parity-dependent energy spectrum, the collapse-and-revival dynamics, and the generation of quadrature squeezed states.

While those fractional revivals suggested the emergence of spatially separated phase-space structures, mapping their structural geometry remained unstudied. We note that geometric separation in the Husimi representation does not establish quantum coherence by itself. Therefore, the aim of this work is to construct the phase-space representations of the Dunkl anharmonic oscillator to visualize this separation. We focus on how the Dunkl parameter $\mu$ modifies the algebraic structure, inducing a parity-dependent redistribution of the phase-space components. This structural effect leads to asymmetric multicomponent wave packets with unequal peak heights.

This paper is organized as follows. In Section 2, we present the algebraic structure and the nonlinear dynamics of the system, defining the time-evolved state and classifying the revival phenomena. In Section 3, we construct the phase-space representations, deriving the Dunkl-Husimi $Q$-function and computing its normalized angular Husimi profile. Section 4 is dedicated to analyzing the phase-space dynamics, focusing on the parity-induced angular anisotropy and the generation of asymmetric multicomponent states. Finally, we present our concluding remarks in Section 5.

\section{Dynamical framework of the Dunkl anharmonic oscillator}

The theoretical model studied in this manuscript relies on the Dunkl derivative, a differential-difference operator that modifies the standard momentum representation by incorporating a reflection symmetry term,
\begin{equation}
D_{\mu} = \frac{d}{dx} + \frac{\mu}{x}(1 - R),
\end{equation}
where $\mu > -1/2$ is the Dunkl parameter and $R$ is the reflection operator \cite{dunkl1989,dunkl2001}. The generalized annihilation and creation operators are defined in terms of the Dunkl derivative as \cite{genest2013}
\begin{equation}
a_{\mu} = \frac{1}{\sqrt{2}}\left(x + D_{\mu}\right), \qquad
a_{\mu}^\dagger = \frac{1}{\sqrt{2}}\left(x - D_{\mu}\right).
\end{equation}
The action of these ladder operators on the orthonormal Dunkl Fock states $|n\rangle$ is given by
\begin{equation}
a_\mu |n\rangle = \sqrt{[n]_\mu} |n-1\rangle, \qquad
a_\mu^\dagger |n\rangle = \sqrt{[n+1]_\mu} |n+1\rangle,
\end{equation}
where the Dunkl integer is defined as $[n]_\mu = n + \mu\left(1 - (-1)^n\right)$.

The Dunkl generalization of the anharmonic oscillator (Kerr medium) is constructed by replacing the standard ladder operators with their deformed counterparts \cite{ojeda2026b}
\begin{equation}
H_{\mu} = \omega a_{\mu}^\dagger a_{\mu} + \frac{\lambda}{2} (a_{\mu}^\dagger)^2 a_{\mu}^2,
\end{equation}
where $\omega$ is the field frequency and $\lambda$ is the anharmonicity constant.

The spectrum, collapse-and-revival dynamics, quadrature evolution, and survival probability of this model were derived in Ref. \cite{ojeda2026b} using the $su(1,1)$ Lie algebra. By defining the generalized number operator $\hat{n}|n\rangle = n|n\rangle$ alongside the reflection operator $R|n\rangle = (-1)^n|n\rangle$, the effective Hamiltonian can be expressed in the Dunkl-Fock basis as
\begin{equation}\label{H_operator}
H_\mu = \frac{\lambda}{2}\hat{n}(\hat{n}-1) + (\omega + \mu\lambda)\hat{n} + \mu\left(\omega - \frac{\lambda}{2}\right)(1 - R).
\end{equation}

The energy spectrum can be written in a single formula valid for any generalized excitation number $n$
\begin{equation}\label{E_n_unified}
E_n^\mu = \frac{\lambda}{2}n(n-1) + (\omega + \mu\lambda)n + \mu\left(\omega - \frac{\lambda}{2}\right)[1 - (-1)^n].
\end{equation}

Since the three components of the Hamiltonian depend on the mutually commuting operators $\hat{n}$ and $R$, the time evolution operator $U_\mu(t) = \exp(-iH_\mu t)$ can be factorized into a product of three independent unitary transformations
\begin{equation}\label{Propagator}
U_\mu(t) = e^{-i\frac{\lambda t}{2}\hat{n}(\hat{n}-1)} e^{-i(\omega+\mu\lambda)t\hat{n}} e^{-i\mu\left(\omega-\frac{\lambda}{2}\right)t(1-R)}.
\end{equation}
This factorization separates the dynamics into three contributions: the nonlinear Kerr term, the global phase-space rotation, and a parity-dependent phase factor controlling the relative phase between the even- and odd-parity sectors.

Based on this factorization, we distinguish between the characteristic times of this periodic evolution. In nonlinear quantum dynamics, the temporal revivals are classified into three physical phenomena:
\begin{enumerate}
    \item Nonlinear revival time ($T_{\text{Kerr}}$). This is the timescale at which the quadratic phase dispersion resets. From the factorized evolution operator in Eq. (\ref{Propagator}), this occurs at $T_{\text{Kerr}}=\frac{2\pi}{\lambda}$, showing that the periodicity associated with the nonlinear Kerr interaction remains invariant. However, the reset of the nonlinear contribution alone does not imply a reconstruction of the quantum state, since the linear and parity-dependent phases must also resynchronize.
    \item Shape revival. A shape revival occurs when, in addition to the reset of the nonlinear Kerr dispersion, the residual linear and parity-dependent phases preserve the wave-packet morphology, possibly up to a phase-space rotation.
    \item State revival. This represents a reconstruction where the wave packet recovers its shape and its initial phase-space position. This condition is satisfied when the survival probability (fidelity) reaches unity, $F(T) = |\langle\psi(0)|\psi(T)\rangle|^2 = 1$ \cite{ojeda2026b}.
\end{enumerate}

As shown in our previous work by computing the overlap amplitude \cite{ojeda2026b}, a state revival requires commensurability conditions among the oscillator parameters so that the nonlinear, linear, and parity-dependent phases resynchronize. Throughout this work, we use the parameter ratio $\omega/\lambda=20$ together with $\mu\in\{0,0.5,1.0\}$. For these parameter values, the phase conditions are satisfied at the nonlinear revival time $t=T_{\text{Kerr}}=\frac{2\pi}{\lambda}$, leading to a reconstruction of the initial state.

To study the phase-space geometry, we define an initial state with a classical-like behavior. We consider the Dunkl coherent state $|\alpha\rangle_\mu$, which is written as a superposition of its even and odd parity components \cite{ghazouani2022,sedaghatnia2023}
\begin{equation}\label{pch}
|\psi(0)\rangle = |\alpha\rangle_\mu = \mathcal{N}_\alpha \left( \sum_{m=0}^\infty \frac{\alpha^{2m}}{\sqrt{[2m]_\mu!}} |2m\rangle + \sum_{m=0}^\infty \frac{\alpha^{2m+1}}{\sqrt{[2m+1]_\mu!}} |2m+1\rangle \right),
\end{equation}
where $[0]_\mu! \equiv 1$, $[n]_\mu! = [n]_\mu[n-1]_\mu\cdots[1]_\mu$ for $n\geq 1$ is the generalized factorial, and the normalization constant is given by
\begin{equation}
\mathcal{N}_{\alpha} = \left[ \sum_{n=0}^\infty \frac{|\alpha|^{2n}}{[n]_\mu!} \right]^{-1/2}.
\end{equation}
We note that the deformation parameter $\mu_G$ used by Ghazouani in Ref. \cite{ghazouani2022} is related to our convention by $\mu_G = \mu - 1/2$.

By applying the time evolution operator to this initial coherent state, the time-evolved state at any time $t$ is obtained as
\begin{equation}\label{PsiT}
|\psi(t)\rangle = \mathcal{N}_\alpha \left( \sum_{m=0}^\infty \frac{\alpha^{2m}}{\sqrt{[2m]_\mu!}} e^{-i E_{2m}^\mu t} |2m\rangle + \sum_{m=0}^\infty \frac{\alpha^{2m+1}}{\sqrt{[2m+1]_\mu!}} e^{-i E_{2m+1}^\mu t} |2m+1\rangle \right).
\end{equation}
This state contains the dynamical information of the Dunkl anharmonic oscillator. In the following sections, this state serves as the foundation to construct the representations and visualize the wave packet geometry in the complex plane.

\section{Phase-space representation: The Dunkl-Husimi $Q$-function}

The phase-space representation of quantum states provides a visualization of the dynamics of nonlinear systems, as well as the phase-space structure of multicomponent and parity-deformed wave packets. In standard quantum optics, the Husimi $Q$-function is defined as the expectation value of the density operator with respect to the coherent state basis \cite{husimi1940,cahill1969,schleich2001,buzek1992}. While the resolution of the identity for Dunkl coherent states has been established \cite{ghazouani2022}, the use of the Dunkl-Husimi representation to characterize the nonlinear phase-space dynamics considered here has not been explored. Therefore, we define the Dunkl-deformed Husimi $Q$-function as the expectation value of the time-evolved density operator $\rho(t) = |\psi(t)\rangle\langle\psi(t)|$ with respect to a test Dunkl coherent state $|\beta\rangle_\mu$,
\begin{equation}\label{Qdef}
Q_\mu(\beta, t) = {}_{\mu}\langle \beta | \rho(t) | \beta \rangle_\mu = |\langle \beta | \psi(t) \rangle_\mu|^2,
\end{equation}
where $\beta = |\beta|e^{i\theta}$ is a complex parameter representing the coordinates in the phase-space.

In order to compute this distribution, we first consider the dual of the Dunkl coherent state introduced in Eq. (8), which is written as
\begin{equation}
{}_{\mu}\langle\beta| = \mathcal{N}_{\beta} \left( \sum_{k=0}^{\infty} \frac{(\beta^{*})^{2k}}{\sqrt{[2k]_{\mu}!}} \langle 2k| + \sum_{k=0}^{\infty} \frac{(\beta^{*})^{2k+1}}{\sqrt{[2k+1]_{\mu}!}} \langle 2k+1| \right),
\end{equation}
where $\mathcal{N}_\beta$ is the corresponding normalization constant. By projecting the time-evolved state $|\psi(t)\rangle$ from Eq. (\ref{PsiT}) onto this basis, the orthogonality of the Dunkl number states, $\langle n | m \rangle_\mu = \delta_{n,m}$, ensures that the overlap amplitude $\mathcal{A}(\beta, t) = \langle \beta | \psi(t) \rangle_\mu$ splits into the two independent parity components
\begin{equation}\label{Overlap}
\mathcal{A}(\beta, t) = \mathcal{N}_\alpha \mathcal{N}_\beta \left( \sum_{m=0}^\infty \frac{(\alpha \beta^*)^{2m}}{[2m]_\mu!} e^{-i E_{2m}^\mu t} + \sum_{m=0}^\infty \frac{(\alpha \beta^*)^{2m+1}}{[2m+1]_\mu!} e^{-i E_{2m+1}^\mu t} \right).
\end{equation}
Notice that this amplitude is a superposition of the even and odd parity contributions, $\mathcal{A}(\beta, t) = \mathcal{N}_\alpha \mathcal{N}_\beta [\mathcal{A}_{even} + \mathcal{A}_{odd}]$. Consequently, the Dunkl-Husimi $Q$-function is given by the squared modulus of this amplitude
\begin{equation}\label{QFinal}
Q_\mu(\beta, t) = \mathcal{N}_\alpha^2 \mathcal{N}_\beta^2 \left| \mathcal{A}_{even}(\beta, t) + \mathcal{A}_{odd}(\beta, t) \right|^2.
\end{equation}
This expression contains a cross contribution between the projected even- and odd-parity amplitudes. The expansion of Eq. (\ref{QFinal}) contains a cross-term proportional to $\text{Re}(\mathcal{A}_{even} \mathcal{A}_{odd}^*)$. While the state $|\psi(t)\rangle$ is pure due to the unitary evolution given by $U(t)$, the Husimi $Q$-function is a non-negative distribution that smooths out oscillatory phase-space modulations. Therefore, rather than visualizing the quantum coherence, our Dunkl-Husimi representation serves to map the phase-space separation and multimodal geometry of the emerging asymmetric states analyzed in Section 4. Finally, it is straightforward to verify from the unified spectrum and the generalized factorials that in the undeformed limit, this distribution reduces to the standard Kerr Husimi $Q$-function,
\begin{equation}
\lim_{\mu \to 0} Q_\mu(\beta, t) = Q_{\text{Kerr}}(\beta, t).
\end{equation}

\subsection{Normalized angular Husimi profile}

While the Dunkl-Husimi $Q$-function provides a three-dimensional geometrical representation of the evolving quantum state, the angular anisotropy can be quantified by analyzing the phase profile of the wave packet. To study the angular distribution of the total Husimi function $Q_\mu(r e^{i\theta}, t)$ in the complex plane $\mathbb{C}$, we define the normalized angular Husimi profile $P_\mu(\theta, t)$ by integrating the Husimi function over the radial measure $r dr$,
\begin{equation}\label{Pintegral}
P_\mu(\theta, t) = \frac{\int_0^\infty Q_\mu(r e^{i\theta}, t) r dr}{\int_{-\pi}^\pi d\theta' \int_0^\infty Q_\mu(r e^{i\theta'}, t) r dr}.
\end{equation}
This definition extracts the angular distribution from the visualized object using the Euclidean area element of the complex $\beta$-plane, guaranteeing that the profile is normalized, $\int_{-\pi}^\pi P_\mu(\theta, t) d\theta = 1$, for any value of the Dunkl parameter. We emphasize that $P_\mu(\theta, t)$ is introduced as a normalized angular descriptor of the total Husimi function and should not be interpreted as a phase probability distribution associated with a phase positive operator-valued measure or with the parity-resolved Segal-Bargmann measure.

To obtain an expansion for this angular profile, we substitute the polar form $\beta=re^{i\theta}$ into the probability amplitude. Noting that the normalization constant $\mathcal{N}_{\beta}$ depends exclusively on the radial coordinate $r=|\beta|$ and the Dunkl parameter $\mu$, i.e., $\mathcal{N}_{\beta} \equiv \mathcal{N}_{\beta}(r,\mu)$, and by expanding the squared modulus of the superposition, the Dunkl-Husimi function can be written as a double sum over the generalized quantum numbers $n$ and $l$
\begin{equation}
Q_{\mu}(r,\theta,t)=\mathcal{N}_{\alpha}^{2}\mathcal{N}_{\beta}^{2}(r,\mu)\sum_{n=0}^{\infty}\sum_{l=0}^{\infty}\frac{\alpha^{n}(\alpha^{*})^{l}r^{n+l}}{[n]_{\mu}![l]_{\mu}!}e^{-i(n-l)\theta}e^{-i(E_{n}^{\mu}-E_{l}^{\mu})t}.
\end{equation}
By substituting this expansion into the radial integral, we obtain the series expansion for the angular profile
\begin{equation}\label{AngularHusimiSeries}
P_\mu(\theta, t) = \frac{\mathcal{N}_\alpha^2}{\mathcal{Z}_\mu} \sum_{n=0}^\infty \sum_{l=0}^\infty \frac{\alpha^n (\alpha^*)^l}{[n]_\mu! [l]_\mu!} \mathcal{R}_\mu(n,l) e^{-i(n-l)\theta} e^{-i(E_n^\mu - E_l^\mu)t},
\end{equation}
where the radial integration factors are given by the standard moments over the complex plane
\begin{equation}
\mathcal{R}_\mu(n,l) = \int_0^\infty \mathcal{N}_\beta^2(r,\mu) \, r^{n+l+1} dr.
\end{equation}
By integrating over the angular variable, the orthogonality condition
\begin{equation}
\int_{-\pi}^\pi e^{-i(n-l)\theta} d\theta = 2\pi\delta_{nl},
\end{equation}
eliminates the time-dependent phases, demonstrating that the global angular normalization factor $\mathcal{Z}_\mu$ is a time-independent constant given by
\begin{equation}
\mathcal{Z}_\mu = 2\pi\mathcal{N}_\alpha^2 \sum_{n=0}^\infty \frac{|\alpha|^{2n}}{([n]_\mu!)^2} \mathcal{R}_\mu(n,n).
\end{equation}
For the numerical results presented below, the angular profile was evaluated from Eq. (\ref{Pintegral}) by numerical radial integration, and the normalization condition was verified to numerical accuracy.

\section{Phase-Space Dynamics and Generation of Asymmetric Multicomponent States}

In this section, we analyze the phase-space dynamics of the Dunkl anharmonic oscillator. We evaluate the Dunkl-Husimi $Q$-function in the complex plane $\beta = \text{Re}(\beta) + i\text{Im}(\beta)$ for various values of the Dunkl parameter $\mu$. For the numerical analysis, we use $\lambda$ as the frequency unit, setting $\lambda=1$ and $\omega/\lambda=20$. The initial coherent amplitude is set to $\alpha = 2.0$, as in Ref. \cite{ojeda2026b}, which yields well-resolved phase-space structures for the dynamics considered here.

The phase-space distributions presented in Figures 1, 2, and 3 are obtained by plotting the analytical expression derived in Eq. (\ref{QFinal}). This is achieved by evaluating the time-dependent probability amplitude $\mathcal{A}(\beta, t)$ given in Eq. (\ref{Overlap}), substituting the parity-dependent energies derived from Eq. (\ref{E_n_unified}). Similarly, the one-dimensional normalized angular Husimi profiles shown in Fig. \ref{fig:pmd} were evaluated by numerical radial integration of the Husimi function for each angular coordinate. Equation (\ref{AngularHusimiSeries}) provides the corresponding analytical series representation. To generate all these numerical surfaces and curves, the infinite series over the generalized excitations are truncated at an integer bound that provides numerical convergence at the precision relevant for the plotted distributions for the selected amplitude $\alpha$. The sums were truncated at $m_{\max} = 20$, corresponding to generalized excitation numbers up to $n = 41$. Numerical stability was checked against larger truncations at representative phase-space points.

\subsection{The physical origin of asymmetry: Structural deformation in the Dunkl formalism}

Before discussing the numerical evaluations of the phase-space dynamics, it is important to understand the physical mechanism that deforms the coherent states. In the standard Kerr medium, the nonlinear phase accumulation is parity-independent because the self-interaction depends only on the excitation number. However, the Dunkl formalism introduces a structural change to the geometry of the system.

When the Hamiltonian is evaluated using its differential realization, the $su(1,1)$ generators carry a singular potential term. Evaluating the quadratic operators from the Dunkl derivative generates an effective parity-dependent inverse-square term proportional to the reflection operator, $V_{\text{eff}} \propto \frac{\mu}{x^2}(1-R)$. This is not an ordinary scalar potential, since its action depends on the parity operator $R$. For the even-parity sector ($R=1$), the term vanishes, whereas for the odd-parity sector ($R=-1$), it produces a repulsive inverse-square contribution proportional to $2\mu/x^2$.

The spatial barrier in coordinate space $x$ is not the primary cause of the generalized operators; rather, both are manifestations of the underlying Dunkl deformation. Consequently, the Dunkl coherent state defined in Eq. (\ref{pch}) incorporates deformed factorials $[n]_\mu!$ that unbalance the probability amplitudes of the even and odd parity components. The resulting morphological deformations of the Husimi $Q$-function in the complex plane $\beta$ are therefore consistent with the same parity-dependent Dunkl structure that, in coordinate space, gives rise to the effective inverse-square term.

Therefore, at $t=0$, before the nonlinear Hamiltonian begins to evolve, the initial coherent state is deformed. The deformation of the relative even- and odd-parity amplitudes produces an asymmetric total Husimi profile at $t = 0$. As the nonlinear Kerr interaction evolves ($t>0$), this structural initial condition contributes to the subsequent phase-space dynamics, transforming the standard symmetric dispersion into an angular anisotropy, generating the asymmetric multicomponent states.

\subsection{Phase-space morphology of the standard Kerr medium ($\mu=0$)}

To quantify the generation of multicomponent wave packets and the angular anisotropy, we analyze the phase-space evolution at five representative times, $t \in \left\{0, \frac{\pi}{8\lambda}, \frac{\pi}{4\lambda}, \frac{\pi}{2\lambda}, \frac{\pi}{\lambda}\right\}$. We present a dual visualization strategy: two-dimensional density maps reveal the phase-space deformation and the locations of the wave-packet components, while three-dimensional surface plots capture the relative peak heights and local Husimi morphology.

\begin{figure*}[htbp]
    \centering
    \begin{subfigure}[b]{0.19\textwidth}
        \centering
        \includegraphics[width=\textwidth]{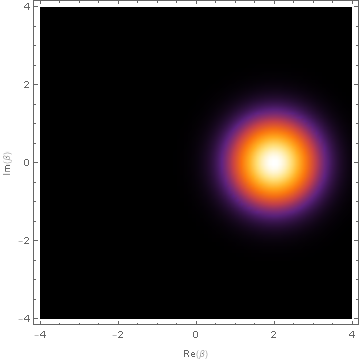}
        \caption{$t=0$}
    \end{subfigure}
    \hfill
    \begin{subfigure}[b]{0.19\textwidth}
        \centering
        \includegraphics[width=\textwidth]{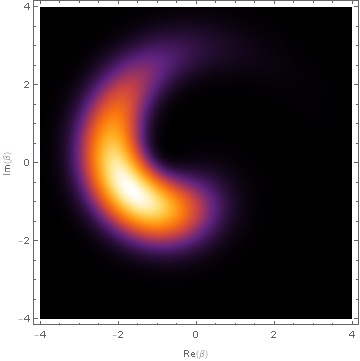}
        \caption{$t=\frac{\pi}{8\lambda}$}
    \end{subfigure}
    \hfill
    \begin{subfigure}[b]{0.19\textwidth}
        \centering
        \includegraphics[width=\textwidth]{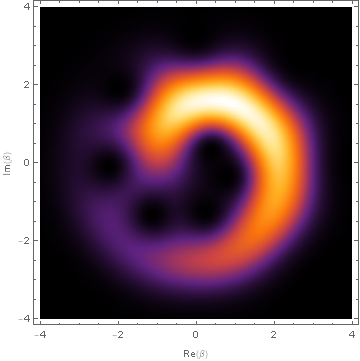}
        \caption{$t=\frac{\pi}{4\lambda}$}
    \end{subfigure}
    \hfill
    \begin{subfigure}[b]{0.19\textwidth}
        \centering
        \includegraphics[width=\textwidth]{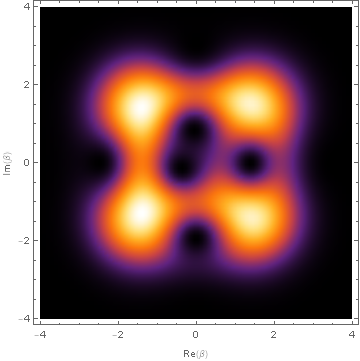}
        \caption{$t=\frac{\pi}{2\lambda}$}
    \end{subfigure}
    \hfill
    \begin{subfigure}[b]{0.19\textwidth}
        \centering
        \includegraphics[width=\textwidth]{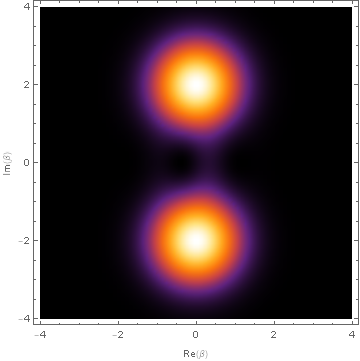}
        \caption{$t=\frac{\pi}{\lambda}$}
    \end{subfigure}

    \vspace{0.3cm} 

    \begin{subfigure}[b]{0.19\textwidth}
        \centering
        \includegraphics[width=\textwidth]{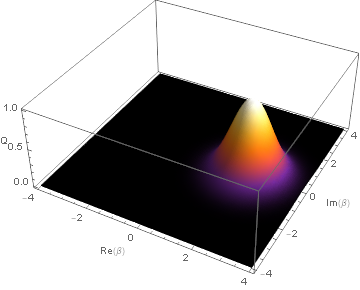}
        \caption*{} 
    \end{subfigure}
    \hfill
    \begin{subfigure}[b]{0.19\textwidth}
        \centering
        \includegraphics[width=\textwidth]{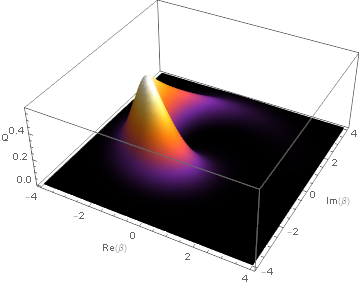}
        \caption*{}
    \end{subfigure}
    \hfill
    \begin{subfigure}[b]{0.19\textwidth}
        \centering
        \includegraphics[width=\textwidth]{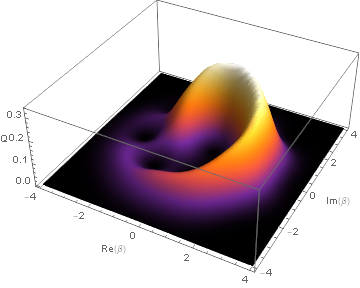}
        \caption*{}
    \end{subfigure}
    \hfill
    \begin{subfigure}[b]{0.19\textwidth}
        \centering
        \includegraphics[width=\textwidth]{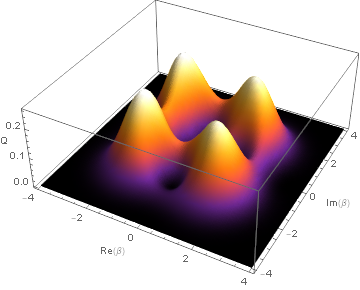}
        \caption*{}
    \end{subfigure}
    \hfill
    \begin{subfigure}[b]{0.19\textwidth}
        \centering
        \includegraphics[width=\textwidth]{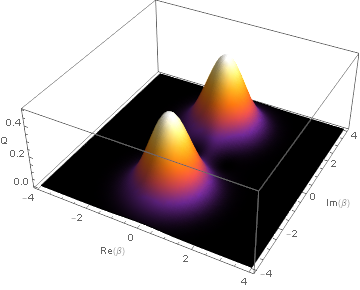}
        \caption*{}
    \end{subfigure}

    \caption{Dual phase-space evolution of the standard Husimi $Q$-function ($\mu=0$). The top row displays the two-dimensional density maps, illustrating the symmetric phase dispersion. The bottom row presents the corresponding three-dimensional geometrical surfaces, confirming the balanced relative Husimi densities of the emerging multicomponent patterns. The sequence captures the initial state ($t=0$), early nonlinear squeezing ($t=\frac{\pi}{8\lambda}$), fractional multicomponent states ($t=\frac{\pi}{4\lambda}$), the symmetric four-component state ($t=\frac{\pi}{2\lambda}$), and the bipartite state ($t=\frac{\pi}{\lambda}$).}
    \label{fig:mu0}
\end{figure*}

To establish a clear reference for the phase-space deformation, we first analyze the standard Kerr medium ($\mu=0$) as a symmetric baseline, where the parity-dependent deformation is absent. The dual visualization of this case is presented in Fig. \ref{fig:mu0}. At the initial time $t = 0$, the standard coherent state appears as a symmetric Gaussian distribution centered on the real axis at $\text{Re}(\beta) \approx 2$. This circular symmetry in both the two-dimensional density map and the three-dimensional surface plot confirms that the phase-space distribution is isotropic around the localized wave packet.

As time advances to $t = \frac{\pi}{8\lambda}$, the nonlinear phase dispersion of the Kerr medium elongates the wave packet. The distribution stretches along the classical circular trajectory of the phase-space, forming a curved arc. As demonstrated by the variance analysis in Ref. \cite{ojeda2026b}, this time lies in the quadrature-squeezing regime.

At $t = \frac{\pi}{4\lambda}$, the wave packet undergoes further phase dispersion, leading to early phase-space modulation. At this stage, the distribution spreads around the circular ring, forming an irregular, inhomogeneous structure. Mathematically, this corresponds to an eight-component fractional superposition; however, due to the smoothing properties of the Husimi $Q$-function, the individual peaks overlap, appearing as a continuous ring with localized density fluctuations.

Morphological features emerge at the fractional revival times. At $t = \frac{\pi}{2\lambda}$, the wave packet fragments into a symmetric four-component state. In the two-dimensional density map (Fig. \ref{fig:mu0}d, top), the components are distributed in a symmetric cross-like shape. Moreover, the corresponding three-dimensional surface plot (Fig. \ref{fig:mu0}d, bottom) confirms that the four peaks exhibit comparable heights, producing an approximately balanced four-component Husimi pattern.

At $t = \frac{\pi}{\lambda}$, the system reaches the half-time revival, collapsing into the bipartite state. The wave packet is split into two identical resolved components diametrically opposed along the imaginary axis and separated by a region of near-zero Husimi density at the origin. As observed in Fig. \ref{fig:mu0}e, the two peaks remain symmetric and equivalent in magnitude.

\subsection{Dynamics of fractional Dunkl deformations ($\mu=0.5$)}

\begin{figure}[H]
    \centering
    \begin{subfigure}[b]{0.19\textwidth}
        \centering
        \includegraphics[width=\textwidth]{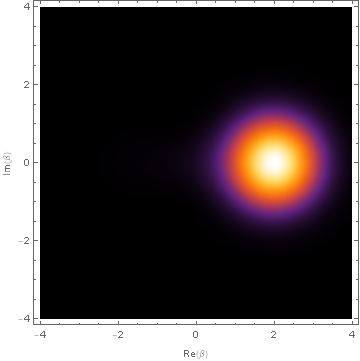}
        \caption{$t=0$}
    \end{subfigure}
    \hfill
    \begin{subfigure}[b]{0.19\textwidth}
        \centering
        \includegraphics[width=\textwidth]{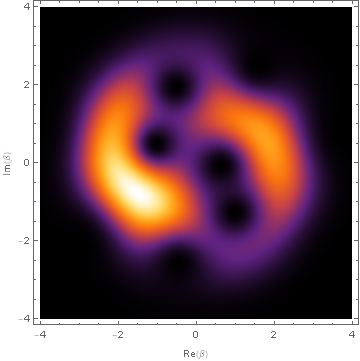}
        \caption{$t=\frac{\pi}{8\lambda}$}
    \end{subfigure}
    \hfill
    \begin{subfigure}[b]{0.19\textwidth}
        \centering
        \includegraphics[width=\textwidth]{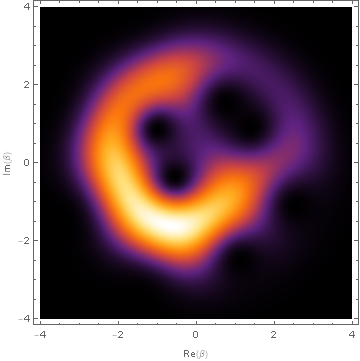}
        \caption{$t=\frac{\pi}{4\lambda}$}
    \end{subfigure}
    \hfill
    \begin{subfigure}[b]{0.19\textwidth}
        \centering
        \includegraphics[width=\textwidth]{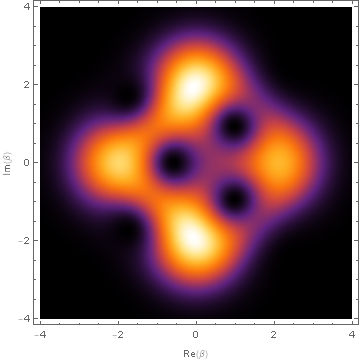}
        \caption{$t=\frac{\pi}{2\lambda}$}
    \end{subfigure}
    \hfill
    \begin{subfigure}[b]{0.19\textwidth}
        \centering
        \includegraphics[width=\textwidth]{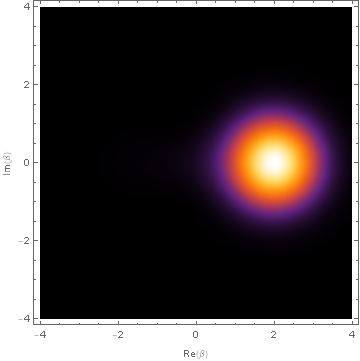}
        \caption{$t=\frac{\pi}{\lambda}$}
    \end{subfigure}

    \vspace{0.3cm} 

    \begin{subfigure}[b]{0.19\textwidth}
        \centering
        \includegraphics[width=\textwidth]{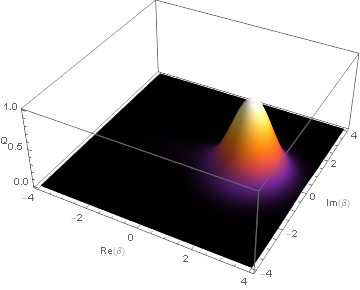}
        \caption*{} 
    \end{subfigure}
    \hfill
    \begin{subfigure}[b]{0.19\textwidth}
        \centering
        \includegraphics[width=\textwidth]{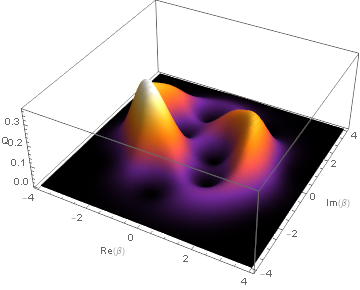}
        \caption*{}
    \end{subfigure}
    \hfill
    \begin{subfigure}[b]{0.19\textwidth}
        \centering
        \includegraphics[width=\textwidth]{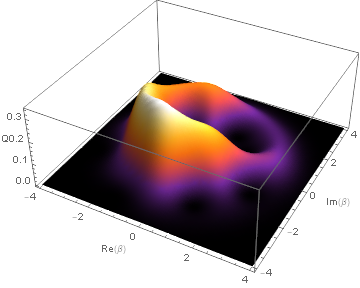}
        \caption*{}
    \end{subfigure}
    \hfill
    \begin{subfigure}[b]{0.19\textwidth}
        \centering
        \includegraphics[width=\textwidth]{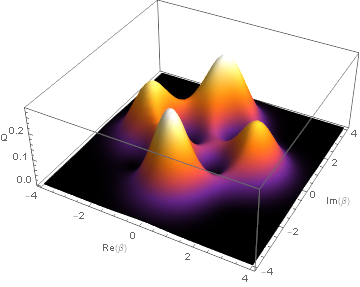}
        \caption*{}
    \end{subfigure}
    \hfill
    \begin{subfigure}[b]{0.19\textwidth}
        \centering
        \includegraphics[width=\textwidth]{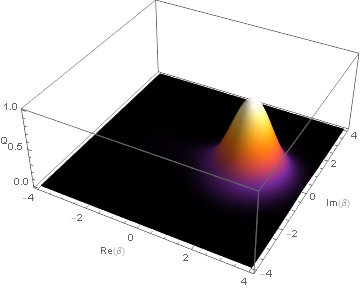}
        \caption*{}
    \end{subfigure}

    \caption{Dual phase-space evolution for the semi-integer Dunkl deformation ($\mu=0.5$). The top row (two-dimensional maps) captures the structural deformation of the initial coherent state at $t=0$ due to the modified statistical weights. The bottom row (three-dimensional surfaces) reveals the unbalanced relative Husimi densities. The sequence demonstrates the parity-induced dynamics leading to early fragmentation ($t=\frac{\pi}{8\lambda}$), the emergence of asymmetric multicomponent states with disparate peak heights ($t=\frac{\pi}{2\lambda}$), and the phase resynchronization resulting in an early single-packet reconstruction at $t=\frac{\pi}{\lambda}$.}
    \label{fig:mu05}
\end{figure}

When the semi-integer Dunkl deformation ($\mu=0.5$) is introduced, the phase-space dynamics undergo a morphological transformation. This transformation reflects both the parity-dependent energy shifts and the underlying Dunkl deformation, whose coordinate-space realization contains an effective inverse-square term. As observed in the initial state ($t=0$) of Fig. \ref{fig:mu05}, the wave packet is no longer a symmetric Gaussian. The initial Dunkl coherent state exhibits an asymmetric tail toward the origin, reflecting the deformation of the coherent-state coefficients relative to the standard case. Furthermore, as the magnitude of the Dunkl parameter $\mu$ increases, this structural phase-space deformation at $t=0$ becomes more perceptible, driving the parity-induced angular anisotropy from the very beginning of the temporal evolution.

The parity-dependent factor in the evolution operator introduces an additional phase in the odd-parity sector, thereby modifying the relative phase between the even and odd components. While the standard Kerr medium only shows a squeezed continuous stretch at $t=\frac{\pi}{8\lambda}$, the parity-dependent relative phase redistributes the Husimi density, producing two resolved components at this early stage. As confirmed by the three-dimensional surface plot (Fig. \ref{fig:mu05}b, bottom), these two prematurely formed components possess unequal peak heights, demonstrating an unbalanced angular density.

As the evolution reaches $t=\frac{\pi}{4\lambda}$, the system exhibits an asymmetric multicomponent pattern. Unlike the approximately ring-like distribution observed for $\mu=0$, the distribution here is displaced, accumulating the majority of the Husimi density on the left side of the phase-space geometry.

At $t=\frac{\pi}{2\lambda}$, the system develops a four-component Husimi pattern whose fourfold symmetry is distorted. Analytically, for the semi-integer deformation $\mu=0.5$ and our commensurate parameters, the parity-dependent linear phase cancels the fractional Kerr rotational shift modulo $2\pi$ for all generalized excitations. This cancellation rotates the pattern by $45^\circ$ compared to the standard medium, positioning the peaks on the coordinate axes. Consequently, we observe an asymmetric four-component Husimi pattern with nonuniform peak heights, which is consistent with parity-induced angular redistribution.

At $t=\frac{\pi}{\lambda}$, the system achieves an earlier reconstruction. While the standard Kerr medium splits into the classic bipartite state at this time, the Dunkl oscillator exhibits a parity-assisted phase resynchronization. Using the factorized evolution operator derived in Eq. (\ref{Propagator}), the total accumulated phase $\Phi_n$ for a Dunkl number state $|n\rangle$ at $t = \frac{\pi}{\lambda}$ for the semi-integer deformation $\mu = 1/2$ is given by
\begin{equation}
\Phi_n = \frac{\pi}{2}n(n-1) + \pi\left(\frac{\omega}{\lambda} + \frac{1}{2}\right)n + \frac{\pi}{4}\left(\frac{2\omega}{\lambda} - 1\right)[1 - (-1)^n].
\end{equation}
To demonstrate the commensurability conditions that make a reconstruction possible, we impose our specific numerical choice $\omega/\lambda = 20$. By discarding all terms that are strict multiples of $2\pi$, the modular arithmetic of the phase simplifies to
\begin{equation}
\Phi_n \equiv \frac{\pi}{2}n^2 - \frac{\pi}{2}n + \frac{\pi}{2}n - \frac{\pi}{4}[1 - (-1)^n] \equiv \frac{\pi}{2}n^2 - \frac{\pi}{4}[1 - (-1)^n] \pmod{2\pi}.
\end{equation}
Evaluating this simplified phase for both parity sectors independently yields
\begin{align}
\text{Even } (n=2m): \quad \Phi_{2m} &\equiv 2\pi m^2 \equiv 0 \pmod{2\pi}, \\
\text{Odd } (n=2m+1): \quad \Phi_{2m+1} &\equiv \frac{\pi}{2}(4m^2+4m+1) - \frac{\pi}{2} = 2\pi m(m+1) \equiv 0 \pmod{2\pi}.
\end{align}
This proves that for all generalized excitations $n$, the combined linear and parity-dependent phases cancel the fractional Kerr dispersion at $t = \frac{\pi}{\lambda}$. Consequently, for the commensurate parameters $\omega/\lambda = 20$ and $\mu = 0.5$, the Husimi distribution reconstructs a single localized packet at $t = \frac{\pi}{\lambda}$, in agreement with the earlier survival-probability analysis. This visualizes the modified parity-phase synchronization caused by commensurability.

\subsection{Phase-space broadening for the deformation $\mu = 1.0$}

For the deformation $\mu = 1.0$, the description of the dynamics must account for the persistence of the multimodal structure. Starting at $t=0$, the two-dimensional density map shows a phase-space deformation, where the initial Husimi profile becomes elongated and asymmetric relative to the standard case.

As the temporal evolution progresses through $t=\frac{\pi}{8\lambda}$ and $t=\frac{\pi}{4\lambda}$, the distribution spreads. The Husimi distribution for $\mu = 1$ exhibits broadening and overlap between neighboring components associated with the modified excitation-number weights and parity-dependent phase accumulation. Nevertheless, the multicomponent angular structure remains visible at the selected fractional-revival times, showing four distinct structures around $t = \frac{\pi}{2\lambda}$ and two structures around $t = \frac{\pi}{\lambda}$.

This phase-space broadening can be physically understood by examining the generalized statistics, where the presence of the Dunkl factorials $[n]_\mu!$ alters the Poissonian distribution of the standard anharmonic oscillator and modifies the relative phase-accumulation conditions. Consequently, the individual components spread and overlap further in the phase-space without destroying their underlying multimodal nature.

We note that the broadening observed in the Husimi representation does not imply decoherence or loss of purity, since the time evolution remains unitary. Because the initial state is pure, the time-evolved density matrix remains pure at all times ($\text{Tr}(\rho(t)^2) = 1$). Although the wave packet appears delocalized at intermediate times, the global purity is preserved and no dissipative decoherence mechanism is present in the model. Consequently, evaluating the distribution at the full fundamental period $t = \frac{2\pi}{\lambda}$ reconstructs the initial asymmetric peak.

\begin{figure*}[htbp]
    \centering
    \begin{subfigure}[b]{0.19\textwidth}
        \centering
        \includegraphics[width=\textwidth]{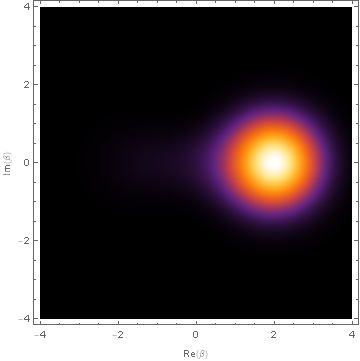}
        \caption{$t=0$}
    \end{subfigure}
    \hfill
    \begin{subfigure}[b]{0.19\textwidth}
        \centering
        \includegraphics[width=\textwidth]{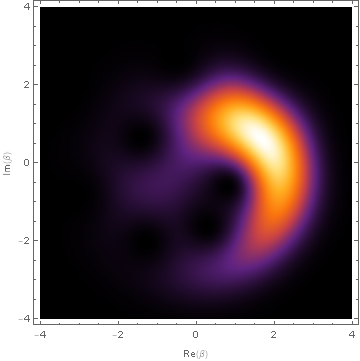}
        \caption{$t=\frac{\pi}{8\lambda}$}
    \end{subfigure}
    \hfill
    \begin{subfigure}[b]{0.19\textwidth}
        \centering
        \includegraphics[width=\textwidth]{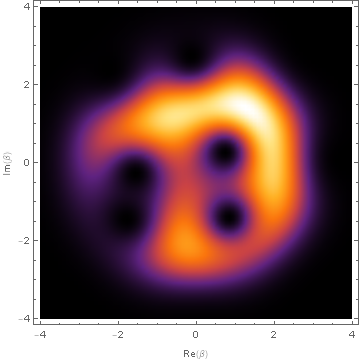}
        \caption{$t=\frac{\pi}{4\lambda}$}
    \end{subfigure}
    \hfill
    \begin{subfigure}[b]{0.19\textwidth}
        \centering
        \includegraphics[width=\textwidth]{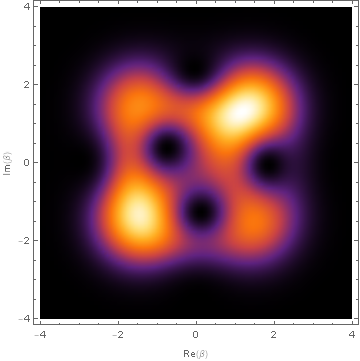}
        \caption{$t=\frac{\pi}{2\lambda}$}
    \end{subfigure}
    \hfill
    \begin{subfigure}[b]{0.19\textwidth}
        \centering
        \includegraphics[width=\textwidth]{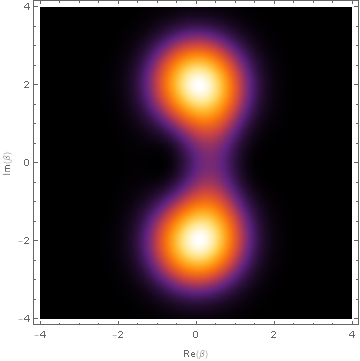}
        \caption{$t=\frac{\pi}{\lambda}$}
    \end{subfigure}

    \vspace{0.3cm} 

    \begin{subfigure}[b]{0.19\textwidth}
        \centering
        \includegraphics[width=\textwidth]{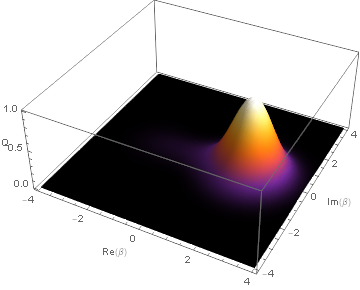}
        \caption*{} 
    \end{subfigure}
    \hfill
    \begin{subfigure}[b]{0.19\textwidth}
        \centering
        \includegraphics[width=\textwidth]{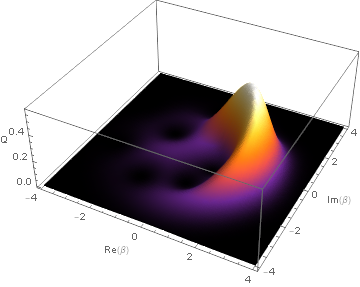}
        \caption*{}
    \end{subfigure}
    \hfill
    \begin{subfigure}[b]{0.19\textwidth}
        \centering
        \includegraphics[width=\textwidth]{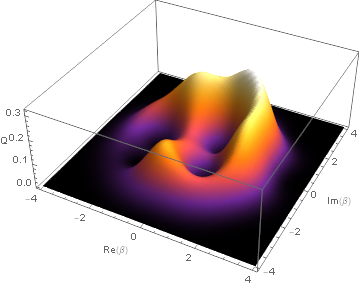}
        \caption*{}
    \end{subfigure}
    \hfill
    \begin{subfigure}[b]{0.19\textwidth}
        \centering
        \includegraphics[width=\textwidth]{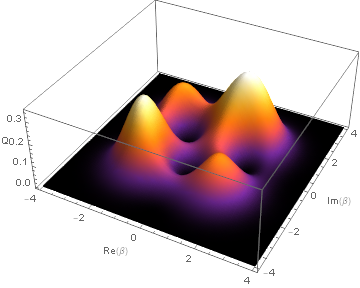}
        \caption*{}
    \end{subfigure}
    \hfill
    \begin{subfigure}[b]{0.19\textwidth}
        \centering
        \includegraphics[width=\textwidth]{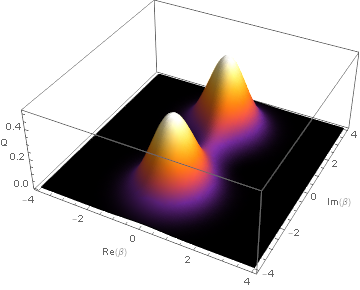}
        \caption*{}
    \end{subfigure}

    \caption{Dual phase-space evolution for the deformation $\mu=1.0$. The deformation produces an elongated and asymmetric initial Husimi profile. At intermediate times, the components broaden and partially overlap while retaining a resolvable multimodal structure.}
    \label{fig:mu1}
\end{figure*}

\subsection{Quantitative angular anisotropy via the normalized angular Husimi profile}

While the three-dimensional surfaces of the Husimi $Q$-function provide a visual representation of the phase-space dynamics, the relative Husimi densities of the multicomponent structures must be quantified to confirm the angular anisotropy. To this end, we evaluate the normalized angular Husimi profile $P_\mu(\theta, t)$, which projects the Husimi function onto the polar angle $\theta \in [-\pi, \pi]$. By tracking the number of local maxima $N_{\text{peaks}}(t)$, their angular positions $\theta_j(t)$, and their respective peak heights $P_{\mu,\max}(t)$, we can substitute visual claims with quantitative descriptors.

\begin{figure}[htbp]
    \centering
    \begin{subfigure}[b]{0.48\textwidth}
        \centering
        \includegraphics[width=\textwidth]{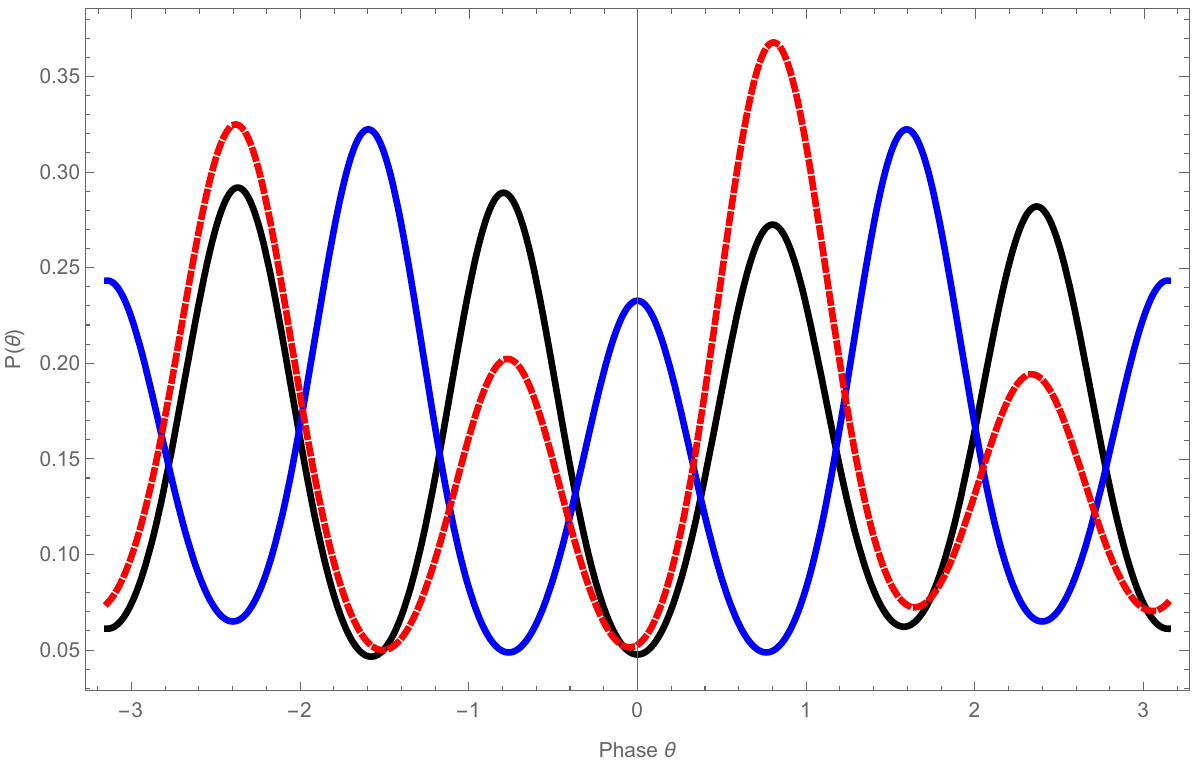}
        \caption{$t=\frac{\pi}{2\lambda}$}
    \end{subfigure}
    \hfill
    \begin{subfigure}[b]{0.48\textwidth}
        \centering
        \includegraphics[width=\textwidth]{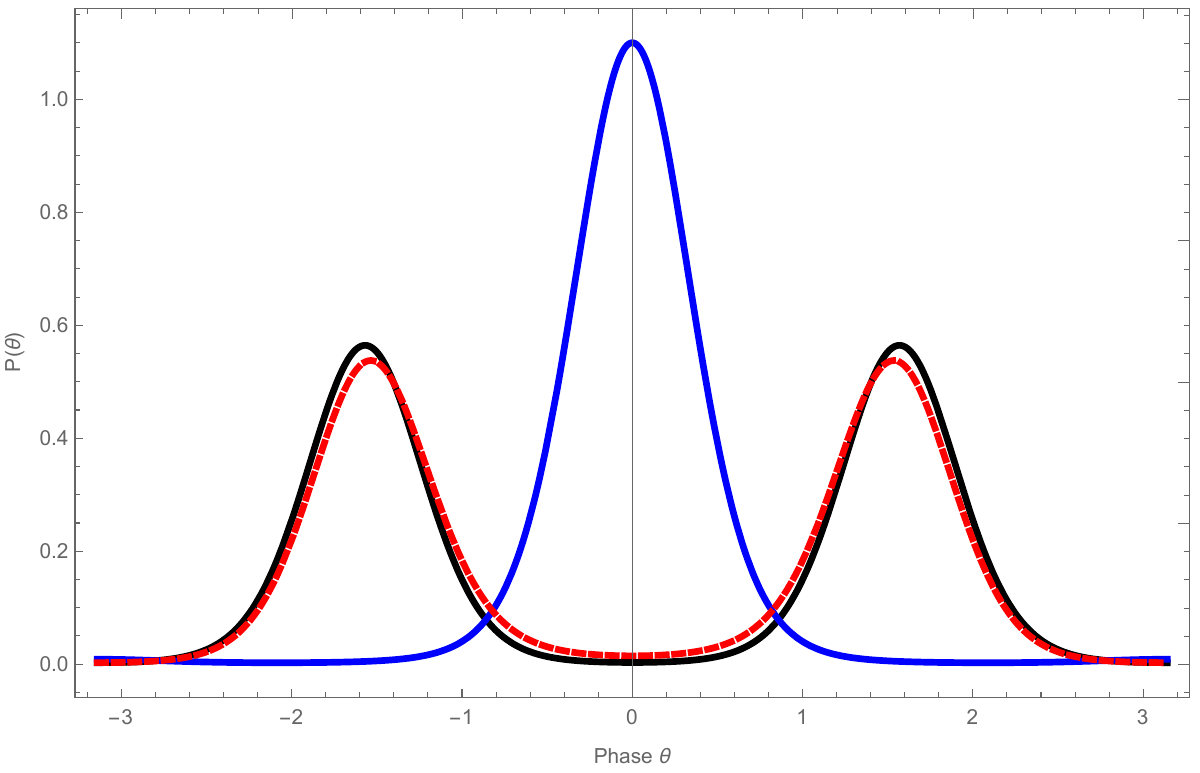}
        \caption{$t=\frac{\pi}{\lambda}$}
    \end{subfigure}
    \caption{Normalized angular Husimi profiles $P_\mu(\theta, t)$ comparing the standard Kerr medium ($\mu=0$, black) with the Dunkl-deformed dynamics for $\mu=0.5$ (blue) and $\mu=1.0$ (red). (a) At $t=\frac{\pi}{2\lambda}$, the nearly balanced four-component profile of the standard case contrasts with the intra-curve asymmetry obtained for $\mu=0.5$, providing quantitative evidence of the angular redistribution associated with the Dunkl deformation. The red curve ($\mu=1.0$) displays broader peaks and elevated Husimi valleys. (b) At $t=\frac{\pi}{\lambda}$, the bipartite state of $\mu=0$ is compared with the resynchronized single-peak reconstruction of $\mu=0.5$ and the broadened bipartite phase of $\mu=1.0$.}
    \label{fig:pmd}
\end{figure}

To evaluate the effect of the Dunkl deformation, we compare the angular profiles for $\mu \in \{0, 0.5, 1.0\}$ at the two characteristic times. Figure \ref{fig:pmd}a displays the profile at $t=\frac{\pi}{2\lambda}$. For the standard Kerr medium ($\mu=0$, black curve), the quantitative features of the symmetric four-component state are confirmed: the profile exhibits $N_{\text{peaks}} = 4$ local maxima positioned at orthogonal angles, forming a nearly balanced four-component angular profile.

However, for the semi-integer deformation ($\mu=0.5$, blue curve), the Dunkl deformation redistributes the relative densities. While the system still displays four local maxima, their heights are different. The endpoints $\theta = -\pi$ and $\theta = \pi$ represent the same angular direction. Since the angular profile is globally normalized, this intra-curve asymmetry provides quantitative evidence of the angular redistribution associated with the Dunkl deformation. Furthermore, for the deformation $\mu = 1.0$ (red curve), the combined modification of the excitation-number weights and parity-dependent phase accumulation produces broader angular profiles; the Husimi density in the angular valleys between the peaks is elevated, providing a quantitative indicator of the broadening of the individual components.

Figure \ref{fig:pmd}b illustrates the angular profiles at half of the nonlinear revival period, $t=\frac{\pi}{\lambda}$. The standard medium ($\mu=0$, black curve) displays the classic bipartite state with two identical peaks. In contrast, the profile for $\mu=0.5$ (blue curve) exhibits a single peak centered at $\theta=0$, providing quantitative evidence of this parity-induced resynchronization: the accumulated phases resynchronize, yielding a single reconstructed wave packet. Finally, for $\mu=1.0$ (red curve), the sum of the linear and parity-dependent phases evaluates to $21\pi n + 19.5\pi[1-(-1)^n]$, which is a multiple of $2\pi$ for both even and odd parity sectors. Consequently, the total accumulated phase matches the standard Kerr quadratic dispersion at this time. Thus, the angular profile remains bipartite and differs only moderately from the standard profile due to the deformed statistical weights.

\section{Concluding remarks}

In this paper, we have investigated the phase-space dynamics and the formation of multicomponent and parity-deformed wave-packet structures in the Dunkl anharmonic oscillator. Using the time-evolved state derived previously in Ref. \cite{ojeda2026b}, which accounts for the parity-dependent energy shifts introduced by the Dunkl deformation, we constructed the phase-space representations of the system.

To characterize the phase-space separation of the wave packet, we derived series expansions for the Dunkl-Husimi $Q$-function and its corresponding normalized angular Husimi profile. Our dual phase-space visualization, which combines two-dimensional density maps and three-dimensional surface plots, demonstrates that the Dunkl parameter $\mu$ modifies the multimodal geometry of the system. Unlike the standard Kerr medium ($\mu=0$), which produces balanced multicomponent states, the Dunkl deformation induces a parity-dependent angular anisotropy. This effect is consistent with the parity-dependent Dunkl structure, whose coordinate-space realization contains an effective inverse-square term.

In the semi-integer regime ($\mu = 0.5$), the parity-dependent deformation leads to the formation of asymmetric multicomponent wave packets, where the individual phase-space components possess unequal peak heights. Furthermore, our phase-space analysis shows that this specific deformation modifies the relative phase between the parity sectors without altering the Kerr nonlinear dispersion, which remains fixed by $\lambda$. Rather than forming a bipartite state at the half-time revival ($t=\frac{\pi}{\lambda}$), the phase shift of the semi-integer Dunkl oscillator acts on the even and odd sectors, resynchronizing their relative phases and reconstructing a single localized wave packet.

These geometrical observations were quantified through the angular Husimi profile. The angular profiles provided quantitative evidence of this parity-dependent redistribution by revealing the unequal heights of the angular-profile peaks. Additionally, for the deformation $\mu = 1.0$, the angular profiles confirmed a broadening associated with the combined modification of the excitation weights and parity-dependent phases. This broadening increases the overlap between neighboring Husimi components, while the unitary evolution preserves the purity of the state.

Despite this broadening, we demonstrated that the system preserves its unitary nature. At the fundamental nonlinear revival time $t=\frac{2\pi}{\lambda}$, the quadratic phase dispersion resets. For the selected parameter ratio, the remaining linear and parity-dependent phases also resynchronize, producing a reconstruction of the wave packet. Finally, we verified that in the limit $\mu \to 0$, our generalized Dunkl-Husimi representation reduces to the standard Kerr Husimi $Q$-function. This study establishes a framework for exploring nonlinear optical processes, asymmetric multicomponent states, and fractional revivals in parity-deformed quantum systems.

\section*{Acknowledgments}

This work was partially supported by SNII-M\'exico, COFAA-IPN, EDI-IPN, and CGPI-IPN Project Number IND-2026-0579.

\section*{Disclosures}

The authors declare no conflicts of interest.

\section*{Data Availability}

No external datasets were used in this study. The numerical data underlying the figures can be reproduced from the analytical expressions and parameters provided in the manuscript.

\end{document}